# Retrospective Comparison of Radiation Oncology Patient Message Responses Generated by Epic and Institutionally Managed LLMs

Meiyun Cao[1, 2], Federico Mastroleo[3, 4, 5], Shuoyang Wei[1], Yuexing Hao[6, 7], Srinivas Seetamsetty[3], Keldon K. Lin[3], Madhurya Reddy[3], Yuzhen Ding[1], Jason Holmes[1], Jonathan E. Moonen[1], Shyanne L. Dobbs[1], Diego Santos Toesca[1], Mark Waddle[3], Wei Liu[1]*.

[1]Department of Radiation Oncology, Mayo Clinic, Phoenix, AZ, United States

[2] Department of Radiation Oncology, University of Kansas Medical Center, Kansas City, KS, United States

[3] Department of Radiation Oncology, Mayo Clinic, Rochester, MN, United States

[4]Division of Radiation Oncology, IEO – European Institute of Oncology IRCCS, Via Ripamonti 435, 20141, Milan, Italy

[5]Department of Oncology and Hemato-Oncology, University of Milan, Via Festa del Perdono 7, 20122 Milan, Italy

[6]Massachusetts Institute of Technology, Cambridge, MA, 02139, USA

[7]MatrAIx, Mountain View, CA, USA

*Corresponding author

## Abstract

We retrospectively compared Epic-generated draft replies with institutionally managed hybrid GPT-4.1 and Gemini 2.5 Pro responses to 51 selected single-turn radiation oncology messages. Six clinicians, blinded to response source, rated overall quality on a five-point scale and perceived editing effort on a four-point scale. Mixed-effects models accounted for repeated ratings by message and clinician. Adjusted overall-quality means were 2.95 for Epic, 4.27 for hybrid GPT-4.1, and 4.39 for hybrid Gemini 2.5 Pro. Corresponding editing-effort means were 2.38, 1.46, and 1.41, with lower scores indicating less perceived effort. Both hybrid models received more favorable ratings than Epic for both outcomes (all Holm-adjusted $P < 0.001$), without a statistically significant difference between the hybrid models. These findings support prospective evaluation of clinician-supervised draft generation, though editing time and patient outcomes were not measured, and clinical safety was not established.

## Introduction

Epic In-Basket (Epic Systems Corporation, Verona, WI, USA) is an electronic messaging workspace within the electronic health record (EHR) that supports asynchronous communication between patients and clinical care teams. It allows clinicians to receive, review, and respond to patient portal inquiries related to symptoms, medications, test results, treatment concerns, appointments, and follow-up care, particularly when real-time or in-person communication is not available[1,2]. However, the volume of In-Basket messages has increased substantially, particularly with the growth of patient portals and changes in communication patterns following the COVID-19 pandemic[1,3]. In radiation oncology, responding to patient inquiries often requires interpretation of treatment history, laboratory and imaging findings, medications, and treatment-related symptoms. Preparing accurate and individualized responses may therefore require clinicians to review information across multiple clinical records[1,4–9]. Timely responses to patient In-Basket messages can support patient safety, continuity and coordination of care, and overall quality of care throughout treatment and follow-up[10,11].

To reduce the growing workload associated with In-Basket messaging, Epic has incorporated generative artificial intelligence(GenAI) into the EHR workflow to generate draft responses to patient messages[4,12–14]. These responses are intended to support clinicians, who remain responsible for reviewing and editing the drafts before they are sent to patients. In recent years, large language models (LLMs) have shown potential for clinical communication tasks including summarization, information retrieval, question answering, and generation of patient-facing responses across multiple healthcare departments, including radiation oncology[2,3,15–20].

Previous studies have compared LLM-generated and clinician-written responses using blinded clinician ratings[2,4,20,21]. Hao et al found comparable ratings for completeness, correctness, and clarity, with higher empathy ratings for LLM responses[2]. Similarly, studies have found comparable information quality and usability but better communication style and perceived empathy for GenAI-generated drafts[2,4]. At the same time, health care organizations are developing institutionally managed LLM environments that can retrieve information from local EHR data and use institution-specific workflows for response generation. However, limited studies are available that have directly compared Epic's commercially integrated LLM-based draft responses with responses generated by institutionally managed LLMs using the same patient inquiries.

In this study, we evaluated patient In-Basket messages across disease sites in Radiation Oncology and compared responses generated by Epic with those produced by two institutionally managed hybrid LLMs. Our primary objective for this study was to compare the overall quality and editing effort of the draft responses generated by Epic and institutionally managed LLMs. Our secondary objectives were to compare linguistic and semantic differences, and multiple quality domains of generated responses. We also sought to assess whether institutionally managed hybrid LLMs could generate high-quality draft responses with reduced clinician-perceived editing effort for clinician-supervised patient portal communication.

## Methods

### Data collection

This retrospective study included a convenience subset of In-Basket messages that require clinical interpretation or judgment from the Radiation Oncology department between November 2025 and March 2026 and was approved by the institutional review board (IRB #: 24-013856). Data collection was conducted in two phases. As shown in **Figure 1**, patient-initiated In-Basket message threads with a corresponding Epic-generated draft response were identified for Phase 1. Messages were categorized by GPT-4.1 into predefined clinical and operational topics and were further filtered to include only single-turn communications consisting of one patient inquiry and one health care worker response. Phase 1 was designed as a focused pilot evaluation of the feasibility and quality of the response-generation approach. To balance evaluation depth with reviewer burden, the 24 most recent eligible messages were selected chronologically for Phase 1 evaluation, prioritizing contemporary messages and minimizing potential differences related to updates in the Epic system. For Phase 2, a separate set of messages collected from January through March 2026 underwent the same initial filtering process. GPT-4.1 was then used to identify messages requiring contextual interpretation or clinical reasoning and to exclude straightforward administrative requests and inquiries requiring access to real-time clinical or operational systems. The screening prompt is provided in ***Supplementary Document S1***. The remaining messages were reviewed by an experienced radiation oncologist with 12 years' experience managing In-Basket messages. We also excluded inquiries that depended on unavailable prior communication or information not accessible to the models, such as attached images or documents. From the eligible Phase 2 messages, 51 were selected chronologically for clinician-blinded evaluation. Not all eligible Phase 2 messages were included in the final evaluation, as the number of messages selected for clinician assessment was defined by the study's structured review design.

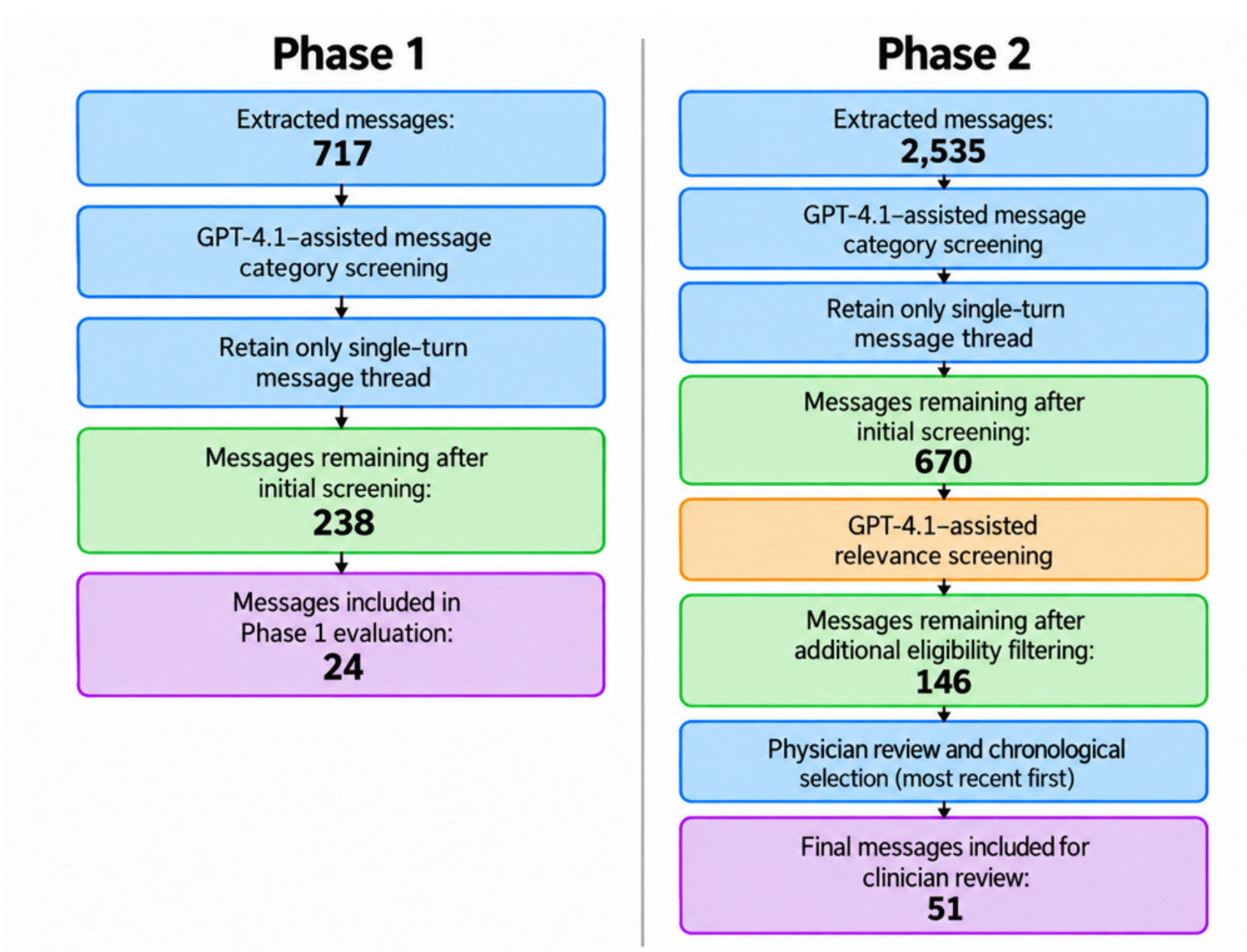

**Figure 1** Workflow for message screening and selection process.

## Evaluation models

Three response-generation models were evaluated: Epic's generative AI–based draft reply model, an institutionally managed hybrid GPT-4.1 model, and an institutionally managed hybrid Gemini 2.5 Pro model. Epic-generated draft replies were integrated within the Epic In-Basket workflow and used vendor-prepared prompts together with selected patient-specific clinical information to generate draft responses[4]. For institutionally managed Hybrid LLMs, Gemini 2.5 Flash was used as a planner model to retrieve clinically relevant patient information, followed by either GPT-4.1 or Gemini 2.5 Pro as a summarizer model for clinical reasoning and final response generation. The planner and summarizer models access the same clinical data resources including the hospital-wide Epic electronic health record (Epic Systems Corporation) and the radiation oncology–specific ARIA database (Varian Medical Systems), and were configured with temperature set to 0, and default top-p parameters to maximize deterministic behavior and minimize variability in retrieval operations.

## Study design

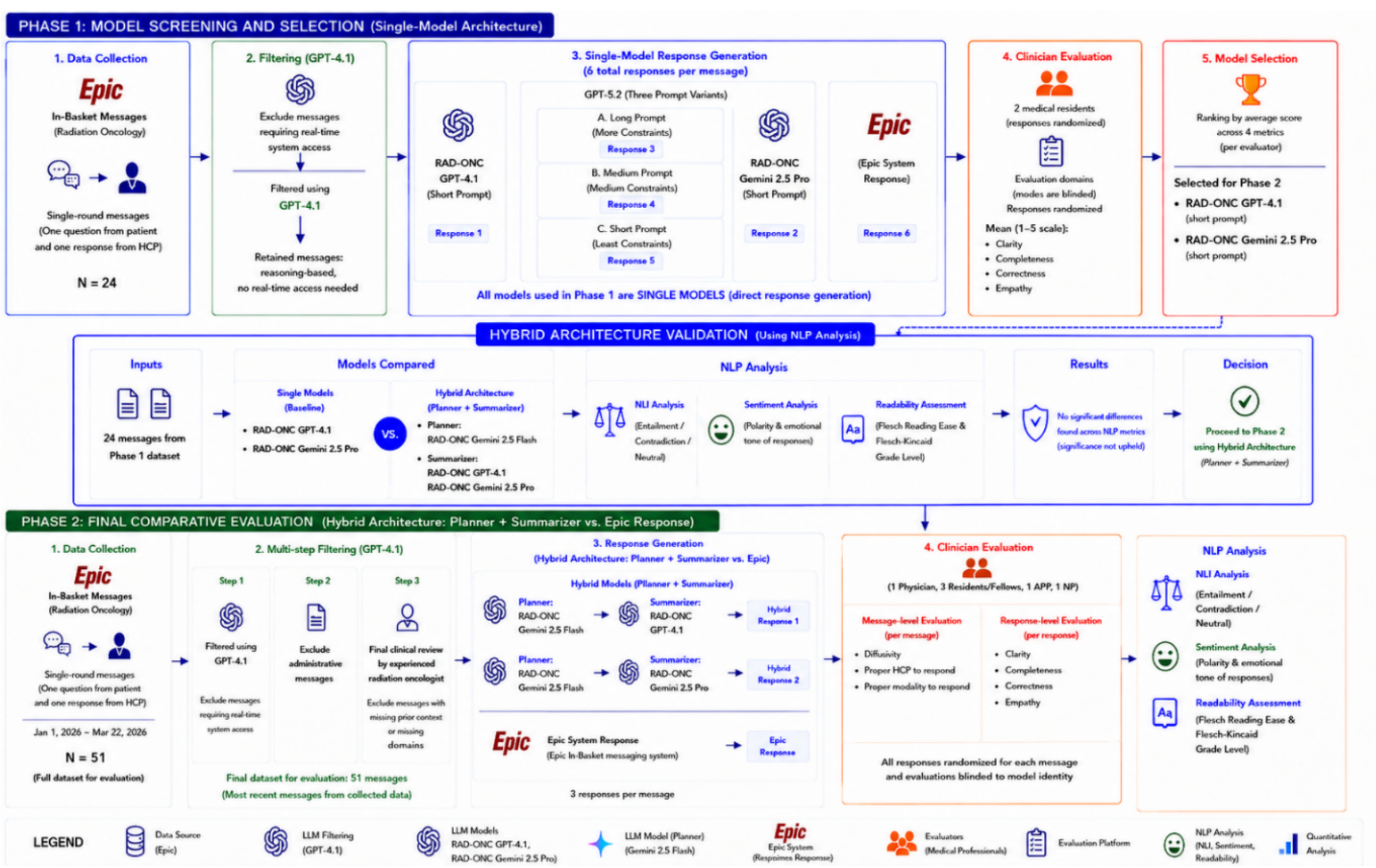

**Figure 2** Two phase study design and evaluation framework for comparing Epic and in-house hybrid LLM responses.

In Phase 1, two medical residents independently evaluated responses generated by different candidate models and by the same model using prompts with different levels of constraint.

Responses were scored for correctness, completeness, clarity, and empathy using a 5-point Likert scale[21]. The average scores across these 4 domains were used to compare model and prompt configurations and to select the LLMs for Phase 2. Detailed Phase 1 evaluation results are provided in ***Supplementary Table S1***.

Before Phase 2, the selected single-model workflows were converted to hybrid architectures to improve token efficiency and separate information retrieval from response generation. Exploratory NLP[2,16,22–24] comparisons using the 24 Phase 1 messages did not identify statistically significant differences between the workflows; however, these analyses were not designed to establish clinical equivalence. Model configurations were finalized before Phase 2 evaluation.

Phase 2 responses were evaluated three AI-generated response systems (Epic, Hybrid GPT-4.1, and Hybrid Gemini 2.5 Pro) through a survey-based assessment. Clinician-authored responses were excluded from the survey but included as a descriptive comparator in the secondary exploratory NLP analysis. The clinician evaluation was conducted in REDCap at both the message and response levels. We recruited six clinicians from Radiation Oncology, representing four distinct clinical roles. During the survey, clinicians had access to patients' medical records as they would when preparing portal message responses in clinical practice. All six clinicians accomplished the survey. At the message level, clinicians assessed response difficulty (5–point Likert scale: 1 indicates least difficult and 5 means most difficult), the most appropriate responder, and the preferred response approach. At the response level, responses from Epic, hybrid GPT-4.1, and hybrid Gemini 2.5 Pro were presented in a randomized order using a single-blind design, with clinician blinded to the response source. Responses were rated on 5-point Likert scales for correctness, completeness, clarity, empathy, and overall quality (1 = least favorable; 5 = most favorable). Perceived editing effort was rated on a 4-point scale (1 = no editing required; 4 = major editing required). Representative prompt example is provided in ***Supplementary document 1***. The Phase 2 evaluation rubrics are provided in ***Supplementary document 2***. Clinicians' characteristics are summarized in Table 1.

**Table 1** Clinician profile

| Clinical role | Department | Gender | Years of practice |
|---|---|---|---|
| Physician | Radiation Oncology | Male | 12 |
| Resident/fellow A | Radiation Oncology | Male | 2 |
| Resident/fellow B | Radiation Oncology | Female | 5 |
| Resident/fellow C | Radiation Oncology | Male | 7 |
| APP | Radiation Oncology | Male | 6 |
| RN | Radiation Oncology | Female | 12 |

**Statistical analysis**

Statistical analyses were conducted using Python, version 3.11.14 (Python Software Foundation), in August 2026. The study design and response evaluation framework were guided by TRIPOD-LLM recommendations[25]. Linear mixed-effects models were used to compare clinician ratings across the 3 response models while accounting for repeated evaluations by clinician and message[29,30]. Pairwise model contrasts were reported as estimated mean differences with 95% confident intervals(CIs), with P values adjusted using the Holm method[30,31]. Message-level Friedman tests followed by Holm-adjusted paired Wilcoxon signed-rank tests were performed as sensitivity analyses of between-model differences [30]. Associations between message difficulty and each outcome were evaluated using linear mixed-effects models with clinician and message as random effects [29,30]. Model-specific difficulty slopes were estimated using Wald tests, and differences in slopes across response models were assessed using difficulty-by-model interaction terms with omnibus Wald $\chi^2$ tests. Pairwise slope differences were evaluated using linear contrasts with Holm-adjusted P values[31]. Interrater reliability across the 6 clinicians was assessed using intraclass correlation coefficients (ICCs) with 95% CIs[32], with pairwise weighted κ coefficients used as a complementary measure of agreement for ordinal ratings[16,33,34]. Exploratory differences across message categories were evaluated using Kruskal-Wallis tests, with Holm adjustment for multiple comparisons[30,31]. All tests were 2-sided, and $P<.05$ was considered statistically significant. As a secondary exploratory analysis, NLP-based analysis[23,24] was used to assess sentiment, readability, response length, and natural language inference (NLI) [26–28] across the three AI-generated response systems. Clinician-authored responses were included as a contextual comparator but were not part of the primary survey evaluation. NLI was performed using BART-large-MNLI (*facebook/bart-large-mnli),* with each patient inquiry as the premise and its corresponding response as the hypothesis.

## Results

### NLI- and NLP-Based Comparison results

NLI analysis showed predominantly high mean neutral probabilities across the three LLMs (**Figure 3A**): 0.9740 for Epic, 0.9810 for Hybrid GPT-4.1, and 0.9711 for Hybrid Gemini 2.5 Pro, compared with 0.7856 for clinician-authored responses. The corresponding mean contradiction probabilities were 0.0253, 0.0131, 0.0247, and 0.1975, respectively. Mean entailment probabilities were low across all four groups (range, 0.0007–0.0169).

NLI probabilities differed significantly across the four matched response groups (Friedman tests, all $P < .001$). Epic demonstrated lower mean entailment probabilities than both hybrid models (Holm-adjusted $P < .001$ for both comparisons), whereas Hybrid GPT-4.1 had a lower mean contradiction probability than Epic (Holm-adjusted $P < .001$). These findings reflect differences in classifier-assigned probabilities rather than clinically verified contradictions.

Sentiment analysis also revealed significant differences across the four matched groups in VADER compound scores (Friedman $\chi^2$[3] = 34.85, P < .001) and TextBlob polarity scores (Friedman $\chi^2$[3] = 10.26, P = .016). Hybrid Gemini 2.5 Pro demonstrated higher TextBlob polarity scores than Epic (Holm-adjusted P = .045) and Hybrid GPT-4.1 (Holm-adjusted P = .018). Additional findings for sentiment, Flesch–Kincaid grade level[16], word count, and sentence count are presented in **Supplementary Figures S1 and S2**.

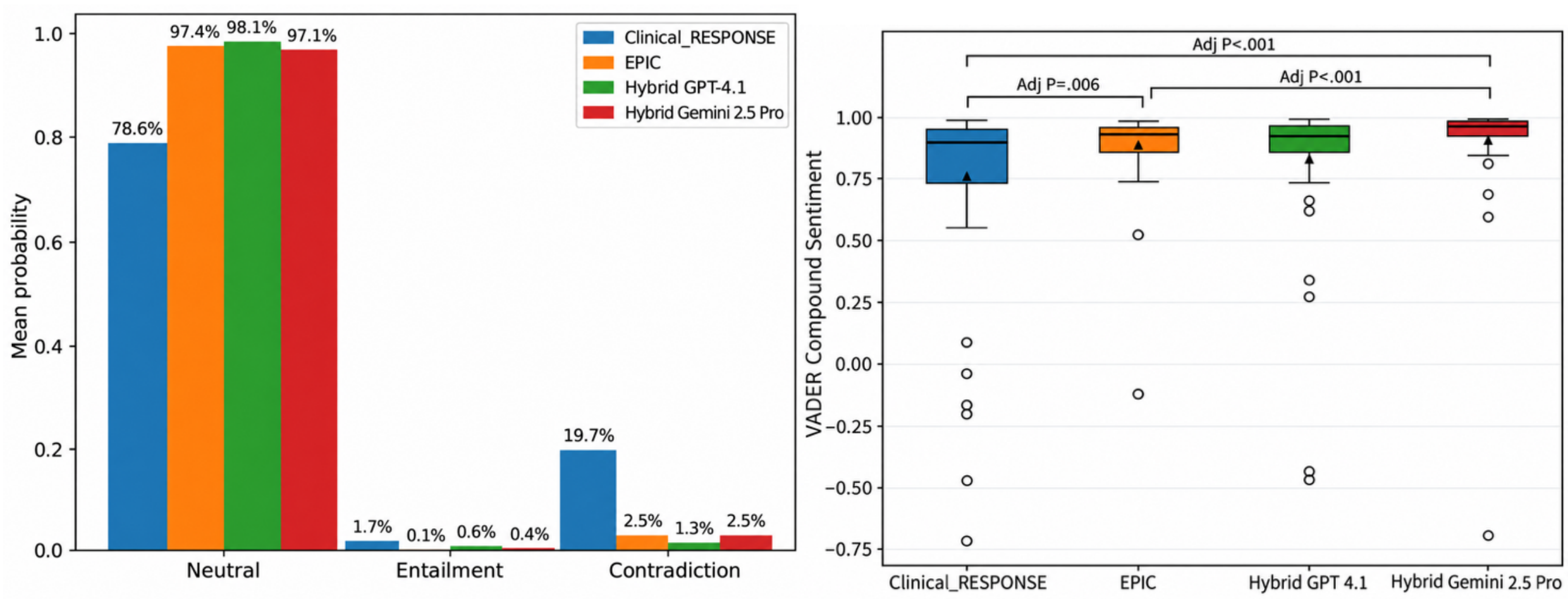

A B

Figure 3 NLP-based comparison of Epic-generated responses, GPT-4.1, and Gemini 2.5 Pro responses. A) Mean natural language inference (NLI) probabilities for neutral, entailment, and contradiction categories across 51 patient inquiry–response pairs per group. Each patient inquiry served as the premise and its corresponding response as the hypothesis. Clinician-authored responses were included as a descriptive comparator, not as a direct semantic reference. Percentages represent mean classifier-assigned probabilities rather than clinically verified classifications. B) Distribution of VADER compound sentiment scores across the four response groups.

## Clinician Evaluation

All six clinicians completed the Phase 2 evaluation, including 1 physician, 3 residents/fellows, 1 advanced practice provider, and 1 registered nurse. The 51 in-basket messages represented 6 clinical categories: radiation treatment questions (n=18), side effects (n=13), medication questions (n=9), medical oncology questions (n=6), laboratory/radiology/pathology reports (n=3), and test results (n=2). Across this evaluation set, the institutionally managed hybrid LLMs received higher clinician ratings than Epic across the assessed response domains.

As shown in **Figure 4**, the overall-quality ratings were higher for both hybrid LLMs than for Epic. In mixed-effects models accounting for repeated evaluations by clinicians and messages, overall-quality scores were 1.33 points higher for Hybrid GPT-4.1 than for Epic (95% CI, 1.18-1.47; Holm-adjusted P<.001) and 1.44 points higher for Hybrid Gemini 2.5 Pro than for Epic (95% CI, 1.30-1.59; Holm-adjusted P<.001). Overall quality did not differ significantly between Hybrid Gemini 2.5 Pro and Hybrid GPT-4.1 (mean difference, 0.11; 95% CI, −0.03 to 0.26; adjusted P=.133) (**Table 2**). Approximately 70% of responses generated by the two hybrid models received clinician-perceived no editing effort score (score = 1: no editing effort), compared with only about 28% of draft responses generated by Epic. In addition, Epic generated drafts received a higher portion of major perceived editing effort score (score = 4: major editing effort) than the two hybrid models. Ratings of clarity, correctness, completeness, and empathy showed a similar overall pattern, with both hybrid LLMs generally rated more favorably than Epic, and the largest descriptive differences were observed for completeness and empathy (***Supplementary Figure S3 and Supplementary Table S2-S3***).

Perceived editing effort was also lower for the hybrid LLMs. Mean editing-effort scores were 2.38 for Epic, 1.46 for Hybrid GPT-4.1, and 1.41 for Hybrid Gemini 2.5 Pro, with lower scores indicating less editing required. Compared with Epic, editing-effort scores were 0.92 points lower for Hybrid GPT-4.1 (95% CI, −1.04 to −0.80; Holm-adjusted P<.001) and 0.96 points lower for Hybrid Gemini 2.5 Pro (95% CI, −1.08 to −0.85; Holm-adjusted P<.001). Perceived editing effort did not differ significantly between the 2 hybrid models (mean difference, −0.05; 95% CI, −0.17 to 0.08; adjusted P=.458). The pattern of lower editing requirements for the hybrid LLMs was generally maintained across clinician roles (**Figure 4**). Both hybrid LLMs were associated with lower perceived editing effort than Epic across clinician groups. Variation in perceived message difficulty across clinicians suggests that difficulty may be more closely associated with clinician role and clinical experience than with the response models (**Figure 5**).

Message-level sensitivity analyses were consistent with the primary mixed-effects results. Friedman tests showed significant differences among response models for both overall quality ($\chi^2$=64.91; P<.001) and perceived editing effort ($\chi^2$=57.43; P<.001). In Holm-adjusted paired Wilcoxon comparisons, both Hybrid GPT-4.1 and Hybrid Gemini 2.5 Pro differed significantly from Epic for both outcomes (all adjusted P<.001), whereas the 2 hybrid models did not differ significantly in overall quality (P=.284) or perceived editing effort (P=.747).

Clinician-rated increasing message difficulty was associated with lower overall-quality ratings and greater perceived for all 3 response models (**Figure 6**). Significant difficulty-by-model interactions were identified for completeness (Wald $\chi^2[2]$ = 20.82; Holm-adjusted P < .001) and overall quality ($\chi^2[2]$ = 9.39; adjusted P = .046), but not for perceived editing effort ($\chi^2[2]$ = 6.31; adjusted P = .171). Epic showed a steeper decline in completeness with increasing clinician-rated difficulty (slope, −0.261; 95% CI, −0.362 to −0.160) than Hybrid GPT-4.1 (−0.042; 95% CI, −0.143 to 0.059) or Hybrid Gemini 2.5 Pro (−0.005; 95% CI, −0.106 to 0.096), with both pairwise differences significant after within-outcome Holm adjustment (P < .001). Detailed analyses of message difficulty by response domain and model are provided in ***Supplementary Table S4 and Supplementary Figures S4-S5***.

No statistically significant differences in response quality or perceived editing effort were detected across message categories after Holm adjustment. These findings can be limited by small subgroup sizes, particularly for test results (n = 2) and laboratory/radiology/pathology reports (n = 3)(***Supplementary Figure S6 and S7***).

Interrater reliability varied across response models and rating domains. For overall quality, ICCs were 0.48 (95% CI, 0.35-0.60) for Epic, 0.35 (95% CI, 0.14-0.53) for Hybrid GPT-4.1, and 0.22 (95% CI, 0.07-0.36) for Hybrid Gemini 2.5 Pro. Mean pairwise weighted κ was 0.53 for overall quality, with higher agreement for empathy (κ=0.61) and completeness (κ=0.58) and lower agreement for clarity (κ=0.36) and correctness (κ=0.25). Itemized interrater agreement analyses are provided in ***Supplementary Figure S8.***

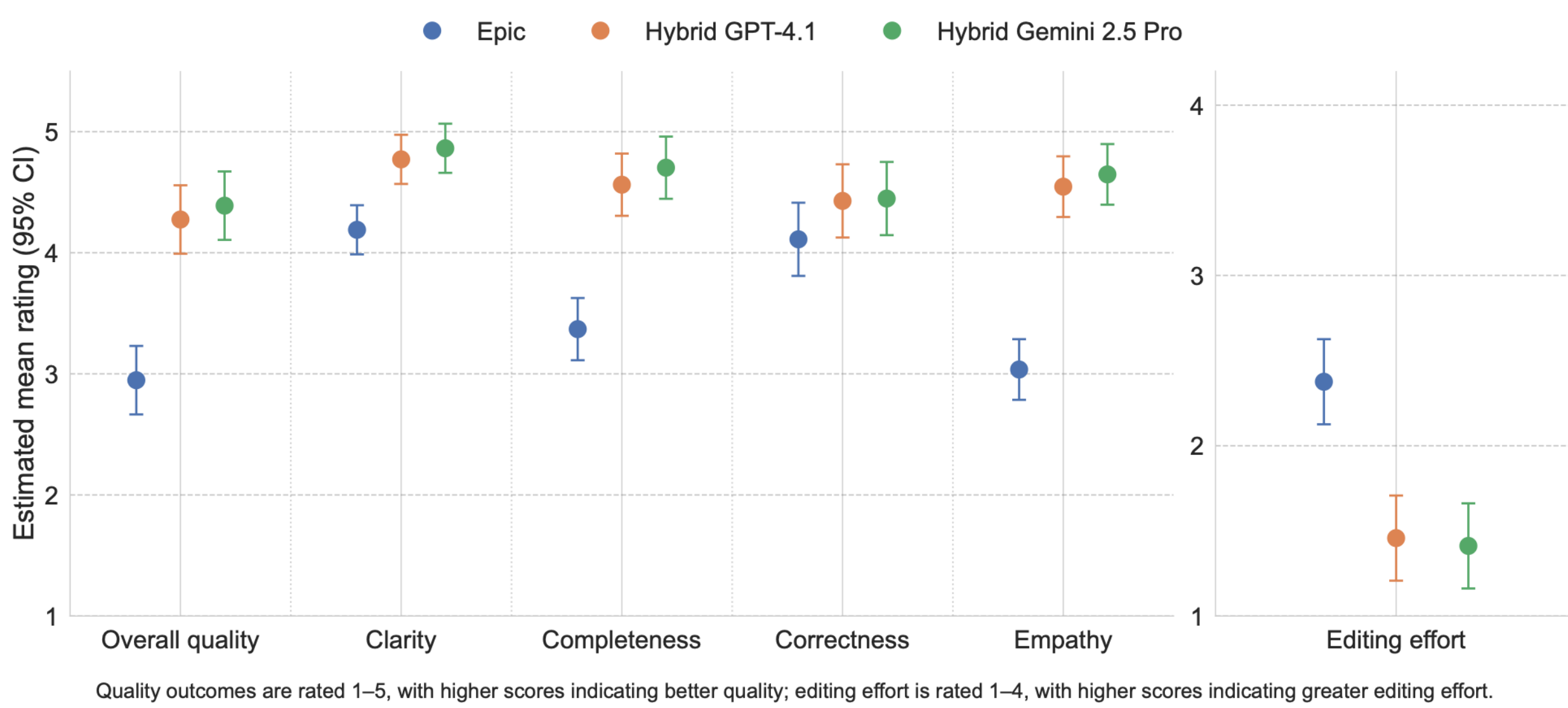

**Figure 4** Mean rating score with 95% CI over response evaluation domains by response models.

**Table 2** Pairwise response models comparison for overall quality and perceived editing effort.

| *Outcome* | *Overall quality* | *perceived Editing effort* |
|---|---|---|
| *Epic, adjusted mean (95% CI)* | 2.95 (2.66 to 3.23) | 2.38 (2.13 to 2.63) |
| *Hybrid GPT-4.1, adjusted mean (95% CI)* | 4.27 (3.99 to 4.56) | 1.46 (1.21 to 1.71) |
| *Hybrid Gemini 2.5 Pro, adjusted mean (95% CI)* | 4.39 (4.11 to 4.67) | 1.41 (1.16 to 1.66) |
| *GPT-4.1 vs Epic, difference (95% CI)* | 1.33 (1.18 to 1.48) | -0.92 (-1.04 to -0.80) |
| *GPT-4.1 vs Epic, Holm-adjusted P* | <.001 | <.001 |
| *Gemini 2.5 Pro vs Epic, difference (95% CI)* | 1.44 (1.29 to 1.59) | -0.96 (-1.08 to -0.84) |
| *Gemini 2.5 Pro vs Epic, Holm-adjusted P* | <.001 | <.001 |
| *Gemini 2.5 Pro vs GPT-4.1, difference (95% CI)* | 0.11 (-0.03 to 0.26) | -0.05 (-0.17 to 0.08) |
| *Gemini 2.5 Pro vs GPT-4.1, Holm-adjusted P* | 0.133 | 0.458 |

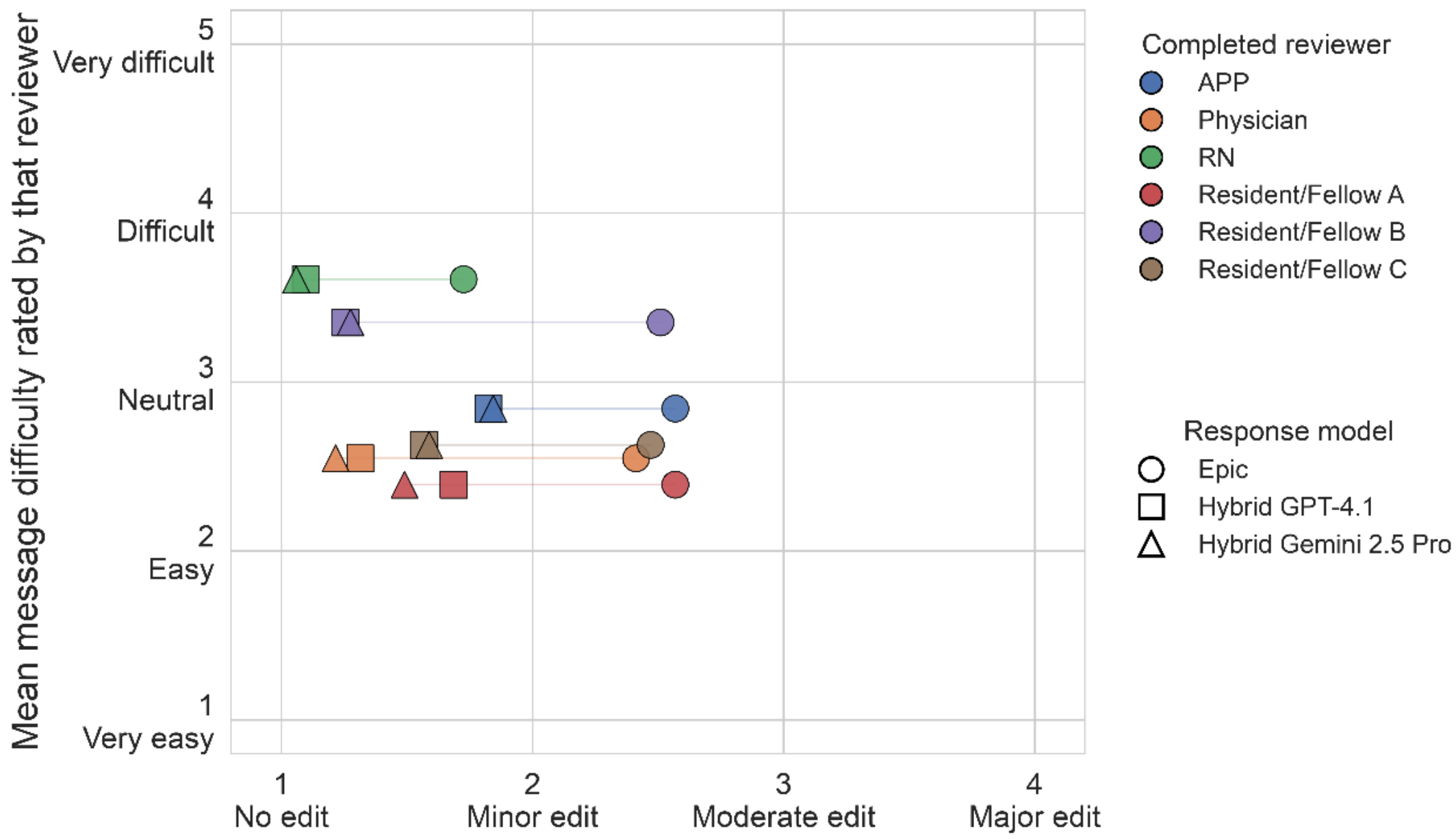

**Figure 5** Relationship between mean message difficulty and perceived editing effort by response model and clinician roles.

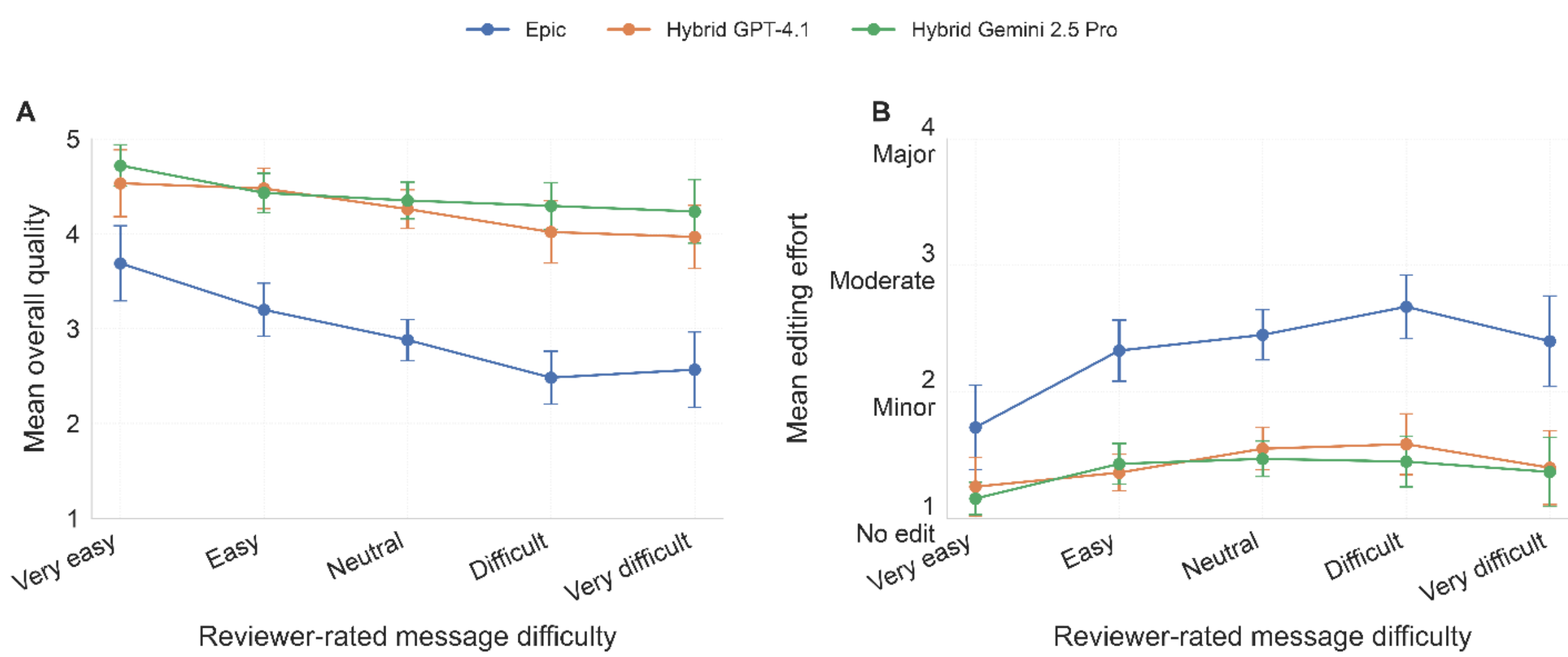

**Figure 6** Message difficulty associated with mean overall quality (A) and mean perceived editing effort(B).

## Discussion

In this retrospective evaluation of 51 radiation oncology patient messages, institutionally managed hybrid models received higher clinician ratings than Epic across all evaluation domains.

The differences were most apparent in completeness and empathy, whereas the differences in correctness were smaller. Increasing message difficulty was generally associated with lower response-quality ratings and greater perceived editing effort, with more pronounced declines in overall quality and completeness for Epic in some comparisons. However, these associations varied among individual clinicians and were not consistently observed across all reviewers. Given the small number of clinicians, some of this variation may reflect differences in clinical judgment and experience. Importantly, lower perceived editing effort does not necessarily indicate reduced clinician workload, as verification of patient-specific information and clinical recommendations may still be required before a response is sent.

The relatively high ratings for completeness and overall quality among the hybrid LLMs may be related to differences in how patient-specific information was retrieved and presented to the response-generation models. The institutionally managed hybrid LLMs retrieved relevant information from multiple EHR and radiation oncology–specific data sources, which was subsequently provided to the response-generation model. In comparison, Epic-generated responses may have had access to a different amount or organization of patient-specific information. The prompts used by the institutionally managed LLMs also contained fewer response constraints, allowing greater flexibility in interpreting the retrieved clinical context and generating patient-facing responses. These differences may have contributed to the higher completeness and overall-quality ratings observed for the hybrid LLMs. However, because the evaluated models differed in architecture, retrieval strategy, available clinical context, prompting, and system-level constraints, the contribution of any individual component could not be determined.

Although the hybrid LLMs generally received higher clinician ratings, clinically important limitations were observed across all three models. Clinicians raised concerns that some recommendations did not adequately reflect patients' individual clinical circumstances. For example, in one prostate cancer case, all three models recommended repeating prostate-specific antigen testing every 6 months, which was not appropriate for that patient's clinical situation. This example illustrates that a recommendation may appear clinically reasonable but remain insufficiently individualized when the patient's treatment history, disease status, or follow-up plan is not adequately considered. Because clinical errors were not systematically assessed across all generated responses, these observations cannot establish error rates or differences in clinical safety between models. Thus, favorable overall-quality ratings do not eliminate the need for clinician verification of patient-specific recommendations.

Interrater reliability varied across response systems and evaluation domains. ICCs indicated low-to-moderate absolute agreement for overall quality, whereas complementary weighted $\kappa$ analyses demonstrated relatively higher agreement for empathy and completeness and lower agreement for clarity and correctness. However, the item-level analysis showed that clinicians frequently differed by no more than 1 point on the 5-point Likert rating scale, particularly for responses generated by the hybrid LLMs. Therefore, relatively low absolute agreement does not necessarily indicate that clinicians had substantially different impressions of a response. These variabilities likely reflect differences in clinical judgment, communication preferences, and expectations regarding appropriate response content. Incorporating multiple clinicians therefore provides a more representative assessment of AI-assisted clinical messaging systems and reduces reliance on the judgment of any single clinician.

The exploratory NLP analysis revealed differences among the Epic and the two hybrid LLMs. Although Hybrid GPT-4.1 demonstrated a lower mean NLI contradiction probability than Epic, the predominance of neutral probabilities may reflect limited discrimination of the classifier in this clinical context and does not establish clinical accuracy or agreement with clinician-authored responses. Similarly, the higher positive sentiment observed with Hybrid Gemini 2.5 Pro does not necessarily indicate greater clinical appropriateness. These findings complement the primary survey-based evaluation.

The purpose of using LLMs for In-Basket response generation is not to replace the clinical care team. Rather, these response models may assist clinicians by preparing an initial response that can be reviewed, edited, and approved before being sent to the patient. Even when a generated response is rated highly, clinicians may still need to verify laboratory values, imaging findings, medications, radiation treatment details, and other patient-specific information. Clinician review is particularly important when the patient inquiry is complex or when the generated response contains a clinical recommendation. Therefore, the potential benefit of these systems is more likely to be reducing drafting effort and supporting clinical communication rather than providing fully autonomous responses.

**Free-text comments review**

In free-text comments, some reviewers noted that Epic-generated responses were less informative and did not directly address the patient’s question. In these cases, the responses often emphasized acknowledgment, reassurance, and supportive language without providing sufficient patient-specific information or clear guidance, which could affect the overall quality of the Epic responses. When we reviewed several of these cases, Epic responses often provided more general information or circled around an uncertain question rather than providing a clear answer. This may represent a more conservative response strategy intended to avoid an unsupported recommendation, but it can also result in a response that is less useful to the patient and requires additional clinician editing. By contrast, the hybrid LLMs more often incorporated retrieved patient-specific information into the response and provided a more direct explanation.

One consideration in this retrospective study is whether the models may have had access to clinical information documented after the original patient inquiry. ne clinician flagged a response to message #10 and a response to message #33 as potentially incorporating information documented after the corresponding patient inquiries. These two messages and their corresponding responses were therefore reviewed by an experienced radiation oncology physician, who examined the source documentation and clinical timeline. The review confirmed that one response to message #10 incorporated information that became available after the patient inquiry was received. For the responses corresponding to message #32, the physician found no apparent evidence that they relied on information documented after the inquiry date. Given the initial concern and the discrepancy between the reviewers' assessments, we conducted a sensitivity analysis excluding both messages and their corresponding responses. This exclusion changed the average scores by no more than 0.02 and did not alter the statistical significance of the study's principal findings, suggesting that the results were robust to the exclusion of these potentially affected cases. However, though the retrieval prompt specifically instructed the models to use only information available before the date of the patient's message, retrospective EHR data may not perfectly reflect the information available to the clinical team at that specific point in time. This highlights an important consideration when retrospectively evaluating LLM systems that retrieve longitudinal EHR data. Future prospective studies could address this issue by preserving a snapshot of the patient record at the time each message is received and restricting model access to information available at that time.

For Gemini 2.5 Pro, despite its strong overall performance, clinician comments also identified instances of unmatched clinical information or incorrect recommendations. For example, one response included medication information that was not documented in the patient's chart, while another advised a patient to fast before a PSA test. These examples are important because even responses that appear clear, complete, and patient-specific may contain clinically relevant errors. Thus, higher overall response quality does not necessarily translate into greater clinical accuracy in every case. Although the clinician comments were not collected systematically and cannot be used to compare error rates across systems, these observations raise important safety consideration for the use of LLM-generated responses in patient communication. Clinician review remains important, particularly when responses include patient-specific recommendations or management guidance.

We also observed differences in clinician preferences regarding how an in-basket message should be managed. For example, the registered nurse frequently commented that some responses should be discussed with the treating physician before being sent and have an average higher difficulty rating than other clinicians. Agreement regarding the most appropriate health care professional to respond to a message also varied among clinicians. These differences may reflect variation in clinical experience, professional roles, practice patterns, and individual preferences regarding the level of certainty required before responding to a patient. Therefore, there may not always be a single correct response strategy for an in-basket inquiry, particularly for messages requiring greater clinical judgment.

## Limitations

This study has several limitations. First, it was a retrospective study conducted within a single radiation oncology department and included 51 selected single-turn messages, which may limit generalizability to other specialties, institutions, message type, and multi-turn conversations. Also, the evaluation was intentionally restricted to messages requiring contextual interpretation or clinical reasoning; using the findings should not be generalized to all patient portal communications, particularly routine administrative requests.

Second, although the physician review found no evidence that evaluated models relied on information entered after the patient inquiry, current retrospective design does not fully exclude this possibility. Future prospective studies could address this limitation by preserving a snapshot of the patient record at the time each in-basket message is received and restricting response generation to information available at that time.

Third, clinician assessment is subjective, and differences in clinical experience, communication preferences, and expectations regarding appropriate response content may influence individual ratings. In addition, there is currently no standardized grading framework for evaluating AI-generated clinical responses. Although correctness, completeness, clarity, and empathy are commonly used evaluation domains, their relative clinical importance may not be equivalent in real-world practice. However, because no validated weighting scheme is currently available, we treated these evaluation domains equally in this study. Future studies may benefit from developing standardized evaluation frameworks that incorporate domain-specific weighting and better reflect the relative clinical importance of different response characteristics.

Fourth, this study evaluated response quality and perceived editing effort rather than prospective clinical outcomes. We did not directly compare clinician workload reduction, patient satisfaction, or whether generated recommendations affected subsequent patient care. The results also reflect the specific model versions, retrieval architecture, prompts, and Epic functionality evaluated during the study period.

Lastly, since the hybrid models and Epic draft responses generative model were different in model architecture, retrieval strategy, available clinical context, prompting, and system constraints, the contribution of individual workflow components could not be isolated.

## Conclusions

In this retrospective evaluation of 51 selected single-turn radiation oncology messages, responses generated by two institutionally managed hybrid LLMs received higher clinician-rated overall-quality scores and lower perceived editing effort than Epic generated draft responses. The largest difference was observed in completeness and empathy, whereas the differences in correctness were smaller. These findings support further evaluation of institutionally managed hybrid LLMs as tools for clinician-supervised patient portal communication. However, variability in clinician ratings and occasional concerns about clinical recommendations indicate that human review remains necessary before responses are sent to patients. Future studies should validate these findings in larger and more diverse clinical settings using prospective study designs and should determine whether improved response quality translates into greater workflow efficiency, reduced clinician workload, and safe patient communication.

**Acknowledgements**

This work was supported by the National Institutes of Health (NIH) through the National Institute of Biomedical Imaging and Bioengineering (NIBIB; grant R01EB293388) and the National Cancer Institute (NCI; grant R01CA280134), as well as Schmidt Sciences.